\documentstyle[multicol,aps,prb]{revtex}

\begin{document}
\draft
\preprint{LA-UR-98-528}

\title{\bf Bosonization in arbitrary dimensions}
\author{Shirish M. Chitanvis}
\address{
Theoretical Division, 
Los Alamos National Laboratory\\
Los Alamos, New Mexico\ \ 87545\\}

\date{\today}
\maketitle
\begin{abstract}
Using methods of functional integration, and performing simple
Gaussian integrals, I show that an interacting system of electrons
can be bosonized in arbitrary dimensions, in terms of the
electrostatic potential which mediates the interaction between them.
Working with the bosonic field,
the sytem is shown to exhibit localized structures 
reminiscent of striping in the cuprates.
\end{abstract}


\begin{multicols}{2}
Attempts at bosonizing a gas of interacting fermions date back to
Tomonoga\cite{tomo}, in the context of superconductivity.
Luttinger\cite{lut} followed with an exact solution of a model
Hamiltonian in one 
dimension.
Lieb and Mattis\cite{lm} perfected Luttinger's solution shortly
thereafter.
Luther\cite{luther} and Haldane\cite{haldane0} published seminal
papers which extended the older one-dimensional methods to higher
dimensions.
Their ideas have been developed in recent years by Houghton et
al\cite{houghton}, Feldman et al\cite{trub}, Castro-Neto and
Fradkin\cite{cnf}, and Frohlich and Gotschmann\cite{froh}. 
Other theoretical methods involving techniques to
bosonize fermions in one or two dimensions have also appeared in the
literature\cite{mloss,hein,kotliar,castro,sei}.
These ideas almost exclusively entail the use of fermionic creation
and annihilation
operators, and as such are difficult to apply in greater than one
spatial dimension.

One of the chief motivations for bosonizing a system of interacting
electrons is to investigate the possibility of these electrons forming
bound states which may not be accessible via the usual perturbative methods.
Recently, a Luttinger liquid has been invoked 
by Anderson\cite{pwa} in an effort to
understand high temperature superconductivity (HTSC).
We will provide in this paper a simpler, exact alternative to the
Luther-Haldane formalism. 
The main focus of this paper will be to establish a straightforward
way to replace a system of interacting electrons with an equivalent
system of interacting bosons using functional
integral methods.
In this paper it will be shown that bosonization of a certain kind can 
always be applied exactly to interacting fermionic systems in
arbitrary dimensions. 
The ideas established in this paper may lead to insight into the
mechanism of high temperature superconductivity, especially as the
detailed physics of the HTSC compounds is included into the problem.
This paper should also be applicable to strongly correlated electron
systems in general.
It should also be relevant to the study of striping in the cuprates.
In fact we will provide later in the paper a qualitative model for
this phenomenon of striping.

Bosonization has attracted 
attention in work on relativistic field theories as well.
In this paper I shall carry out in the non-relativistic regime
appropriate to condensed matter physics, the program of bosonization developed
recently by Bannerjee in relativistic Quantum Field Theory
(QFT)\cite{banner1,banner2}. 
Bannerjee used the method of functional integration to eliminate
matter fields in the standard expression for the action, leaving
just bosonic fields for consideration.
The focus in this paper will be on phenomena which emerge
when these methods are applied to issues in condensed matter physics
for d $\ge 1$.

Our technique is related closely to the work of Fogedby\cite{fog}, Lee and
Chen\cite{lee} and Kopietz\cite{kop}. 
The focus in these papers has been the computation of the electronic
Green's function, in an effort to understand the origin of
non-Fermi-liquid behavior.
I shall instead concentrate on uncovering physics which lies in the
effective action for the bosonic fields that replace the fermionic
fields.
As indicated above, I obtain a tentative model for striping in the
cuprates.
This paper shows how to eliminate the approximations invoked by
Kopietz to carry out successfully the program of bosonization in $d > 1$.
Last but not least, the techniques utlizied here will be seen to be
more economical than in previous papers, as a result of which the final, novel
results are more transparent.
Rather than diminishing previous results, it is hoped that the
perspectives of this paper can be combined with existing techniques to 
shed more light on HTSC.

Without further ado, I note that the usual partition function
for a gas of interacting electrons (in atomic units; $\hbar = c = e =
m = 1$) is given by

\begin{eqnarray}
Q =&& \int {\cal D} \psi^*  {\cal D}\psi \exp{iS} \nonumber\\
S =&& \int dt \int d^d x \psi^*(\vec x,t)\left( i\partial_t -H_0\right)
  \psi(\vec x,t) \nonumber\\ 
      -&&\int dt \int d^d x \psi^*(\vec x,t) \int d^d x' \vert
      \psi(\vec x',t)\vert^2  
         \vert\vec x -\vec x' \vert^{-1} \psi(\vec x,t) \nonumber\\
H_0 =&& -{1\over 2}\nabla^2
\label{Q1}
\end{eqnarray}

where $\psi$ and $\psi^*$ are two-vectors describing fermion fields
having a spin 1/2.
$H_0$ can be modified with an extra term describing the interaction of 
the electrons with an underlying lattice.
This partition function can be derived 
by introducing the scalar electrostatic potential:

\begin{eqnarray}
Q \sim&& \int {\cal D} \psi^* {\cal D}\psi {\cal D} \phi~ \exp{iS'} \nonumber\\
S' =&&\int dt \int d^d x \left[\psi^* \left(i \partial_t - \phi\right) \psi -
      \psi^* H_0 \psi -  {\vert\vec \nabla \phi\vert^2\over {8 \pi}} \right]
\label{Q2}
\end{eqnarray}

where an irrelevant normalization constant has been ignored.

Integrating out the electrostatic potential using Gaussian integration 
in \ref{Q2} leads directly to \ref{Q1}.
On the other hand, if the matter fields $\psi$ and
$\psi^*$ are integrated out in \ref{Q2}, the  partition function
involves just the electrostatic field $\phi$.
In general, the electromagnetic vector potential $\vec A$ should also
be included.
Traditionally this is not done in condensed matter physics on
grounds that it is the Coulomb interaction which dominates the
physics.
Elimination of the matter fields then leads to:

\begin{eqnarray}
Q \sim \int {\cal D}\phi \exp &&i \int dt \int d^d x 
         \left\{-  {\vert \vec \nabla \phi\vert^2 \over {8 \pi}
             }\right\} \times \nonumber\\ 
          &&\left(\rm{det}\left[-i\left(i\partial_t -\phi + \mu
                 -H\right)\right] \right)^{-1}
\label{Qeff}
\end{eqnarray}

where $\mu$ signifies a constraint used
to guarantee number conservation, since the integration over the
matter fields includes cases involving broken symmetry.
Note that this constraint is formally equivalent to an interaction of
the electrons with a uniform background field.
The way $\mu$ has been defined, $ \mu > 0$ implies an
interaction with a uniform positive background.
We see here an analogy with the Wigner-Seitz model.
An increasing number density of
electrons is signalled by an increasing $\mu$.
The appearance of the scalar field $\phi$ in \ref{Qeff} implies that
we are dealing with a spin-zero object, but it should be remembered
that $\phi$ is an approximation to the photonic field, which carries a 
unit spin.
Given that the gauge degree of freedom represented by $\phi$ can be
absorbed into the electronic wavefunction, we see that there is a
certain similarity between our approach and that of
Haldane\cite{haldane0} who performed a decomposition 
of the electronic wavefunction in terms of its phase.
I will now evaluate the determinant in \ref{Qeff}, which contains
non-trivial physics.
This evaluation is more convenient in Euclidean space, which entails
setting 
$t \to -i t$.
I then need the determinant of the operator
$\left(\partial_t + \phi - \mu +H_0\right)$.
Upon considering an associated {\it heat diffusion equation} viz.,
$-\partial_{\tau}g\left(\vec x - \vec x',t-t',\tau\right) = 
\left(\partial_t + \phi - \mu +H_0\right) g\left(\vec x - \vec
  x',t-t',\tau\right) $
the determinant is obtained as $\exp-\zeta'(0)$\cite{ramond}, where

\begin{equation}
\zeta(s) = {1\over\Gamma(s)} \int d\tau \tau^{s-1} \int dt\int d^3 x~
g\left( \vec 0,0,\tau \right)
\label{zeta}
\end{equation}

Here attention has been restricted to three dimensions for ease of
presentation.
Our method of evaluating determinants of operators is fairly
well-established\cite{ramond}.
The basic idea is to perform the evaluation formally with a $\phi$
which is constant, and then to use the {\it form} obtained thusly to
write down the complete expression for the effective potential.
The manipulations of this equation are
fairly straightforward, and the only 
care one needs to exercise is in taking Laplace
transforms, considering only those frequency domains which ensure convergence
of the overall expression.
The effective potential which arises
from the determinant is given by:

\begin{eqnarray}
V_{eff}(\phi-\mu) =&& (\vert\phi-\mu\vert)^{5/2} \times \nonumber\\
  &&\left[
     \Psi(-5/2) - \Gamma(-5/2)
                 \ln (\vert\phi-\mu\vert)\right]
\label{eff}
\end{eqnarray}

where $\Psi$ in \ref{eff} is a digamma function and should not be
confused the fermionic field used earlier.
Given that we have here a nonrelativistic formulation of electron
dynamics, it is seen that temporal derivatives do not appear in
the bosonized version of the partition function.
$\phi$ refers to the (time-independent) electrostatic potential.
As such the partition function involving the field $\phi$ can be
redefined with a
delta-function, in much the same way as gauge constraints are imposed
on path-integral formulations.
On classical grounds only $\phi >0$ needs to be considered,
since we have a system of interacting like charges.
More generally, if we are considering quantum effects, negative values
of $\phi$ could occur. 
It can be shown graphically that for sufficiently small values
of $\mu$, the effective potential displays a single
minimum for $\phi > 0$.
There is of course a second minimum for negative values of the field.
I will revisit this double-well form of $V_{eff}$ shortly.
The compact form for $V_{eff}$ was obtained relatively economically.
Besides the ease of derivation, the simple form for $V_{eff}$ displays 
a double-well structure which will shortly yield some novel insights
into the interacting electron system we started out with. 

If the potential is expanded around the positive minimum which occurs for
sufficiently low values of the chemical potential, then the
Euler-Lagrange equation associated with the effective Lagrangian can
be written as: 

\begin{eqnarray}
\left(\nabla^2 -\lambda^{-2}\right)\delta \phi &&= 0\nonumber\\
\lambda^{-2} &&= 4 \pi V''_{eff}(\phi_{min})
\label{ins}
\end{eqnarray}

This is the analog of Laplace's equation for a system interacting via
a {\it screened} Coulomb interaction.
The right hand side of \ref{ins} signifies that the field
configuration is such that there are no free charges in the system.
I interpret this as an insulating phase!
The reason for restricting attention to positive values of the field
is that the Euler-Lagrange equation represents the classical limit.
The basic assumption behind this argument is that the system settles
into a state desribed by $\phi_{min}$ and harmonic perturbations around
it.

More complicated situations than the one described by \ref{ins} can
also occur.
The previous discussion regarding the insulating
regime can be expanded by considering the following double-well
representation for the effective potential, viz.,
$V_{eff} = V_o -(\alpha/2) (\phi-\mu)^2  +  (\beta/4) (\phi-\mu)^4$.
The two minima for this function occur at 
$\phi_{min} = \mu \pm \sqrt{\alpha/\beta}$.
From this it follows that the Euler equation is:

\begin{equation}
\nabla_{\vec \xi}^2 \tilde \phi + \tilde \phi -
\left({\beta\over\alpha}\right) \tilde \phi^3 = 0
\label{soli}
\end{equation}

where $\tilde \phi = \phi -\mu$, and $\vec \xi = \sqrt{4 \pi \alpha} ~\vec x$.
Note that for $\mu$ sufficiently small, there is only one mimimum for
$\phi > 0$.
In one dimension, as $x \to 0$, $\phi \approx (1/2) x^2 (-\mu
+(\beta/\alpha) \mu^3)$, while for $x \to \pm \infty$, $\phi \sim
\phi_{min}$.
This allows us to connect neighboring insulating regions via a domain wall. 
Going to a cylindrical geometry, it is possible to show
approximately that as the 
cylindrical co-ordinate $\rho \to 0$, $\phi-\mu \sim \rho^s$, with
$s$=0,1,2...
While for $\rho\to\infty$, $\phi-\mu \sim \phi_{min} \left[1-s^2/(2
  \rho^2)\right]$. 
By the arguments given above, the domain wall between the defect and
the surrounding insulating phase has free charges flowing through it.
An array of such localized defects are reminiscent of the striping
which occurs in the cuprates.
The model we just described is complementary to the approach of
Castro-Neto et al\cite{castro}.
In spherical geometry, as the radial co-ordinate $r\to 0$,
$\phi-\mu \sim r^l$, with $l$=0,1,2,...
For distances far from the center of such spherulitic defects, $\phi-\mu \sim
 \phi_{min} \left[1-l(l+1)/(2 r^2)\right]$.
This approximate treatment in non-planar geometry can be trivially
generalized using the unabridged form of $V_{eff}$. 

As $\mu$ increases, both minima in $V_{eff}$ occur for positive values 
of the field $\phi$.
As indicated below \ref{Qeff}, the appearance of $\mu$ is formally
equivalent to introducing a Wigner-Seitz positive uniform background
model.
We would therefore like to argue that for sufficiently large values of 
$\mu$, the system behaves as a metal.
We currently cannot describe this transition from an insulating to a
metallic phase.
The only remark we can make is the following:
Given that classically the system can settle into one of two energy
minima, it must be possible to have spatial domains in which the
system has $\phi = \pm \phi_{min}$, with domain walls permitting a
transition from one state to the other.
For example, if we assume that the three-dimensional system exhibits
variation in only one direction, then the domain wall can be described 
by $\phi = \mu + \phi_{min} ~tanh[x \sqrt{2 \pi \alpha}]$.
This domain wall has free charges flowing through it, i.e., it
satisfies \ref{ins}, but with a source term on the right hand side.
But this is similar to the defects discussed above, in the insulating
phase.
It is only if there are a large number of these domain walls, and they form
a connected network, will the system conduct appreciably.

We are currently in the process of computing the electronic
Green's function, given that the system of interacting electrons can
be bosonized.

I would like to acknowledge electronic correspondence with
P.W. Anderson and P. Kopietz rectifying gaps in my
knowledge of previous literature.





\end{multicols}

\end{document}